\documentstyle[12pt]{article}

\catcode`\@=11
\long\def\@makefntext#1{
\protect\noindent \hbox to 3.2pt {\hskip-.9pt
$^{{\ninerm\@thefnmark}}$\hfil}#1\hfill}                

\def\@makefnmark{\hbox to 0pt{$^{\@thefnmark}$\hss}}  

\def\ps@myheadings{\let\@mkboth\@gobbletwo
\def\@oddhead{\hbox{}
\rightmark\hfil\ninerm\thepage}
\def\@oddfoot{}\def\@evenhead{\ninerm\thepage\hfil
\leftmark\hbox{}}\def\@evenfoot{}
\def\sectionmark##1{}\def\subsectionmark##1{}}

\renewcommand{\thefootnote}{\fnsymbol{footnote}}

\def\sectionc{\@startsection {section}{1}{\z@}{-3.5ex plus -1ex minus 
    -.2ex}{2.3ex plus .2ex}{\bf }}
\def\subsectionc{\@startsection{subsection}{2}{\z@}{-3.25ex plus -1ex minus 
   -.2ex}{1.5ex plus .2ex}{\it }}
\renewcommand{\section}[1]{\sectionc{#1}\hspace*{\parindent}}
\renewcommand{\subsection}[1]{\subsectionc{#1}\hspace*{\parindent}}
\newcounter{appendixc}
\newcounter{subappendixc}[appendixc]
\newcounter{subsubappendixc}[subappendixc]

\renewcommand{\appendix}[1] {\vspace*{0.6cm}
        \refstepcounter{appendixc}
        \setcounter{figure}{0}
        \setcounter{table}{0}
        \setcounter{equation}{0}
        \renewcommand{\thefigure}{\Alph{appendixc}.\arabic{figure}}
        \renewcommand{\thetable}{\Alph{appendixc}.\arabic{table}}
        \renewcommand{\theappendixc}{\Alph{appendixc}}
        \renewcommand{\theequation}{\Alph{appendixc}.\arabic{equation}}
        \noindent{\bf Appendix \theappendixc #1}\par\vspace*{0.4cm}}

\def\abstracts#1{{
        \centering{\begin{minipage}{13.2truecm}\footnotesize\baselineskip=13pt\noindent
        \parindent=0pt #1
        \end{minipage}}\par}}

\renewenvironment{thebibliography}[1]
        {\begin{list}{\arabic{enumi}.}
        {\usecounter{enumi}\setlength{\parsep}{0pt}
\setlength{\leftmargin 0.75cm}{\rightmargin 0pt}
         \setlength{\itemsep}{0pt} \settowidth
        {\labelwidth}{#1.}\sloppy}}{\end{list}}

\newcounter{itemlistc}
\newcounter{romanlistc}
\newcounter{alphlistc}
\newcounter{arabiclistc}

\newcommand{\fcaption}[1]{
        \refstepcounter{figure}
        \setbox\@tempboxa = \hbox{\footnotesize Figure~\thefigure. #1}
        \ifdim \wd\@tempboxa > 6in
           {\begin{center}
        \parbox{6in}{\footnotesize\baselineskip=13pt Figure~\thefigure. #1}
            \end{center}}
        \else
             {\begin{center}
             {\footnotesize Figure~\thefigure. #1}
              \end{center}}
        \fi}

\newcommand{\tcaption}[1]{
        \refstepcounter{table}
        \setbox\@tempboxa = \hbox{\footnotesize Table~\thetable. #1}
        \ifdim \wd\@tempboxa > 6in
           {\begin{center}
        \parbox{6in}{\footnotesize\baselineskip=13pt Table~\thetable. #1}
            \end{center}}
        \else
             {\begin{center}
             {\footnotesize Table~\thetable. #1}
              \end{center}}
        \fi}

\def\@citex[#1]#2{\if@filesw\immediate\write\@auxout
        {\string\citation{#2}}\fi
\def\@citea{}\@cite{\@for\@citeb:=#2\do
        {\@citea\def\@citea{,}\@ifundefined
        {b@\@citeb}{{\bf ?}\@warning
        {Citation `\@citeb' on page \thepage \space undefined}}
        {\csname b@\@citeb\endcsname}}}{#1}}

\newif\if@cghi
\def\cite{\@cghitrue\@ifnextchar [{\@tempswatrue
        \@citex}{\@tempswafalse\@citex[]}}
\def\citelow{\@cghifalse\@ifnextchar [{\@tempswatrue
        \@citex}{\@tempswafalse\@citex[]}}
\def\@cite#1#2{{$\null^{#1}$\if@tempswa\typeout
        {IJCGA warning: optional citation argument
        ignored: `#2'} \fi}}

 1
 1
 1

\font\ninerm=cmr9

\newcommand{\nc}{\newcommand}
\nc{\be}{\begin{equation}}
\nc{\ee}{\end{equation}}
\nc{\bea}{\begin{eqnarray}}
\nc{\eea}{\end{eqnarray}}
\nc{\ra}{\rightarrow}
\nc{\M}{{\cal M}}
\nc{\bb}{\bibitem}
\begin{document}
\begin{flushright}
{\footnotesize
Invited talk at the Int. Symposium on Non-Nucleonic \\
Degrees of Freedom Detected in the Nucleus \\
Osaka, Japan, Sept. 2-5, 1996. \\
ADP-96-33/T232
}
\end{flushright}
\vspace*{0.5truein}
\centerline{\normalsize\bf CVC IN PARTICLE PHYSICS}
\baselineskip=15pt
%

\vspace*{0.6cm}
\centerline{\footnotesize ANTHONY W. THOMAS$^1$}
\baselineskip=13pt
\centerline{\footnotesize\it $^1$Department of Physics and Mathematical Physics 
and Institute for Theoretical Physics}
\baselineskip=13pt
\centerline{\footnotesize\it University of Adelaide, 
Adelaide, SA 5005, Australia}
\centerline{\footnotesize E-mail: athomas@physics.adelaide.edu.au}
\vspace*{0.3cm}

\vspace*{0.6cm}
\abstracts{
We review the hypothesis of the conserved vector
current (CVC) within the Standard Model. In addition to 
the classic tests, such as pion beta decay and neutrino
scattering, we mention recent tests involving LEP data.
As well as providing a clear indication that the
isovector current is not conserved, rho-omega mixing
offers a fascinating opportunity to study CP violation
at B-factories and we outline these ideas.
Finally we briefly touch on a new approach to mass
generation in the Standard Model which, for example,
leads to the up-down mass difference which breaks CVC.
}

\normalsize\baselineskip=15pt
\setcounter{footnote}{0}
\renewcommand{\thefootnote}{\alph{footnote}}

\section{Introduction}\label{sec:intro}  
The CVC hypothesis \cite{FG} arose at the time when only charged
current weak interactions were known.  The Hamiltonian
for semi-leptonic, weak interactions was written as
\begin{equation}
H^{SL}_{CC} = \frac{G}{\sqrt{2}} \left[ J^+_{\lambda} L^{\lambda} + 
J^-_{\lambda} L^{\lambda \dagger} \right] ,
\label{eq:1}
\end{equation}
with $J^{\pm}_{\lambda}$ the charge changing hadronic current and 
$L^{\lambda}$
the leptonic current.  The hypothesis had two parts:
firstly, if we break the hadronic weak current into vector and axial
pieces, $J^{\pm}_{\lambda} = V^{\pm}_{\lambda} - A^{\pm}_{\lambda}$, 
that the vector pieces $V^{\pm}_{\lambda}/cos \theta_C$ 
and the isovector piece of the
electromagnetic current were the three components,
$j_{i \lambda}$, of a vector in isospace; secondly, that all 3
components of this current were conserved, $\partial ^{\lambda} j_{i
\lambda} = 0$.
Viewed in hindsight, with the full 
Standard Model at hand, it is hard to appreciate the
power and insight that it represented.

For the purposes of this brief review we begin with
the Standard Model.  The first part of the CVC
hypothesis is then trivial because all interactions
involving strongly interacting systems are built from
the same vector (and axial vector) quark currents.  On
the other hand, the second part of the hypothesis is
incorrect because the QCD Hamiltonian contains a piece
proportional to 
$(m_u - m_d) (\bar{u}u - \bar{d}d)$.  
The fact that $m_u \neq m_d$ implies that  
$\partial ^{\lambda} j_{i \lambda} \propto (m_u - m_d) \neq 0 $ and thus
CVC can only be, at best, a good approximation.
Nevertheless, it has proven such a successful
approximation that it is built into all 
phenomenological treatments.  In the next section we
review two of the classic applications of CVC as well
as an important illustration of the fact that $m_u$ is
not exactly equal to $m_d$.

\section{Classic Tests}\label{sec:class}
The most famous testing grounds for CVC are $\beta$-decay of the free
neutron, bound nucleons and the pion. For the nucleon the most accurate
measurements by far are made in finite nuclei and these are
traditionally not considered part of particle physics. Therefore, we
begin with the case of pion $\beta$-decay, which still gives 
the most precise, particle physics test. We then recall the potential
importance of neutrino scattering where, as we shall see, the errors are
still very large. To complete the section, we briefly review the most
spectacular example of the effect of having $m_u \neq m_d$, namely $\rho
-\omega$ mixing. This phenomenon is also very important in modern tests
of CVC (e.g., at LEP) 
as well as for determinations of $CP$-violation at $B$-factories
-- as we shall discuss in the next section.

\subsection{Pion Beta Decay}\label{subsec:pion}
The decay
\begin{equation}
\pi^- \rightarrow \pi^0 + e^- + \bar{\nu_e},
\label{eq:2}
\end{equation}
is severely depressed by the small mass difference
between the charged and neutral pions.  Only the
vector part of the hadronic weak current can play a
role in this decay.  The CVC hypothesis then naturally
relates the isovector, vector matrix element to the electromagnetic form
factor of the pion ($F_{\pi}$):
\begin{eqnarray}
< \pi^0 (k') | J^{\lambda} | \pi^- (k) > & = & - cos \theta_C \sqrt{2} 
< \pi^- (k') | j_{e-m}^{\lambda} | \pi^- (k) > \nonumber \\
 & = & - cos \theta_C \sqrt{2} F_{\pi}(q^2) (k + k')^{\lambda},
\label{eq}
\end{eqnarray}
with $q = k' - k$.

The very small mass difference involved,
$\Delta \sim 4$ MeV, means that the difference between
$F_{\pi}(0) = 1$ and $F_{\pi}(q^2)$ is negligible.
Hence the $\beta$-decay lifetime can be written in
terms of essentially kinematic quantities:
\begin{equation}
\tau^{-1} = G^2 \frac{cos\theta_C^2}{30\pi^3}
\left[ 1 - \frac{\Delta}{2m} \right] \Delta^5 \bar{F} (1 +
\delta_{\pi}).
\end{equation}
The phase space factor, $\bar{F}$, is near one and the
(electromagnetic) radiative correction ($\delta_{\pi}$)
is of order 1\% \cite{sirlin}.  This leads to a
theoretical branching ratio $1.0482 \pm 0.0048 \times
10^{-8}$ \cite{mcf}.

The most recent experiment to accurately determine
this ratio was carried out at LAMPF more than 10 years
ago.  McFarlane et al. obtained 1224 $\pm$ 36 good events
for $\pi^+ \beta$-decay after observing more than
$10^{11}$ pions \cite{mcf}.  Their final branching
ratio, $1.026 \pm 0.039 \times 10^{-8}$, dominates the
current world average ($1.025 \pm 0.034 \times
10^{-8}$)\cite{pdg}.  Until this time the best
measurement was from Depommier et al.\cite{dep}, in
1968.  Clearly the theoretical and experimental values
are completely consistent at the current experimental
limit of about 3\%.

\subsection{Neutrino Nucleon Scattering}\label{subsec:neutrino}
Neutrino nucleon scattering is the primary source of information on the
weak axial vector current, notably $G_A(q^2)$. However, in order to
extract information on the axial vector current it is more or less
compulsory to assume that the vector matrix elements can be taken from
electromagnetic interactions using CVC. Because of the difficulties of
dealing with a neutrino beam it is preferable to leave a charged nucleon
in the final state. Thus one is led to 
a deuteron target 
because of the lack of a free neutron
target. However, by using the resolving power of a bubble chamber to
tag a low momentum, spectator proton
\begin{equation}
\nu_{\mu} + d \rightarrow \mu^- + p + p_{spectator},
\end{equation}
one can reduce the experimental uncertainties.

The data is not sufficiently accurate to determine the detailed shape of
$G_A(q^2)$, which is usually parametrized as a dipole 
by analogy with the vector form factors. 
It is possible to relax the CVC constraint on the vector form factors
and search simultaneously for the best-fit dipole masses $M_A$ {\it and}
$M_V$. The values obtained in the early 80's by 
Baker {\it et al.} ($M_V = 0.86 \pm
0.07, M_A = 1.04 \pm 0.14$) \cite{baker} and Miller {\it et al.} 
($M_V = 0.96 \pm 0.04,  M_A = 0.80 \pm 0.10$) \cite{miller}
were consistent with CVC in that
$M_V$ measured in electron scattering is 0.84 GeV. 
As there is now very good agreement on $M_A$, at about 1.02 GeV, the
latter value is probably nothing to worry about, even though the
apparent discrepancy in $M_V$ is $2\frac{1}{2}$ standard deviations. The latest
result from the Brookhaven group (Kitagaki {\it et al.}) is reassuring,
yielding
($M_V = 0.89 +0.04 -0.07$ and $M_A = 0.97 +0.14 -0.11$) \cite{kitagaki}.

The same data can also be used to search for evidence for the second-class
vector current. The most recent limit comes from a measurement of the
reaction $\bar{\nu}_{\mu}p \ra \mu^+n$ at Brookhaven \cite{Ahrens}. 
Unfortunately the sensitivity to the
scalar form factor is reduced by a factor $(m_{\mu}/M_N)^2$.
If it is parametrized as 
\begin{equation}
F_S(q^2) = \rho \frac{F_V(0)}{1 - \frac{q^2}{M_S^2}},
\end{equation}
then the limit on $\rho$ is $\rho < 1.8$, with $M_S = 1$GeV and the
axial tensor term set to zero. One can actually set a better limit on
the axial tensor, second class current
but that is not our concern.

\subsection{Neutrino Deep Inelastic Scattering}\label{subsec:neutdis}
Although it is not derived from CVC alone, the Adler sum-rule
provides a fundamental test of the quark currents within the
Standard Model.  It relies on the equal-time commutation relation
\begin{equation}
\delta (x_0) \left[ J^-_0(x),J^+_0(0) \right] = -4\delta^{(4)}(x)
\left[ V^3_0 + A^3_0 \right].
\end{equation}
Taking the matrix element of this relation between hadronic states
and averaging over spin (so that $<A^3_0> = 0$) we find
\begin{equation}
\int^{\infty}_0 d\nu
\left[ W^{\nu}_2(Q^2,\nu) - W^{\bar{\nu}}_2(Q^2,\nu) \right] = 
-4<I_3 >,
\end{equation}
and finally, changing the integration variable to Bjorken $x$, and
using the fact that $\nu W^{\nu}_2 = F^{\nu}_2(x,Q^2)$ scales, we
find
\begin{equation}
\int^1_0 dx \frac{F^{\nu n}_2 - F^{\nu p}_2}{2x} = 1.
\end{equation}
Note that this sum-rule is protected against $0(\alpha_s)$
corrections.  Even now, the best experimental test of this
fundamental sum-rule comes from the long extinct BEBC facility at
CERN, with the result $1.01 \pm 0.08 (stat.) \pm 0.18 (syst.)$ 
\cite{Allas}.  While the data is clearly consistent with expectations,
an error of more than 20\% is really unacceptable in such a
fundamental quantity.

\subsection{Rho-Omega Mixing}\label{subsec:rhom}
The most spectacular evidence that the vector current is not conserved
comes from electromagnetic interactions. In particular, the data for $e^+
e^- \rightarrow \pi^+ \pi^-$, which is dominated by the
isovector $\rho$-meson, shows a sharp interference pattern near the
mass of the isoscalar $\omega$-meson. The natural explanation of this is
that, because $m_u \neq m_d$, the isospin pure $\rho$ and $\omega$ mesons
are not eigenstates of the full QCD Hamiltonian. (N.B. One must, of
course, include the mixing induced by coupling to the photon, namely
$\rho \rightarrow \gamma \rightarrow \omega$, but this is only 10\% of
the observed amplitude. Also, the $\rho$ and $\omega$ are resonances and
therefore not eigenstates of any Hamiltonian, but one can give the
statement some rigour within models, such as the cloudy bag \cite{cbm},
by turning off the coupling to decay channels.)

Suppose we make the standard simplification\cite{Renard},
which is that the direct decay of the isospin pure $\omega$ to
two pions cancels the imaginary piece of the two pion loop
contribution to the mixing self-energy.  
This means that we can neglect
the pure isospin state, $\omega_I$, coupling to two pions
($\M^{\nu}_{\omega_I\ra\pi\pi}=0$) with the understanding that
it is the real part of the mixing amplitude that is being extracted.
To lowest order in the mixing amplitude,
the amplitude for the virtual $\gamma$ to decay to two pions 
can be written:
\begin{equation}
\M^{\mu}_{\gamma\ra\pi\pi} =
\M^{\mu}_{\rho_I\ra\pi\pi} \frac{1}{s_{\rho}} \M_{\gamma\ra\rho_I}
 + \M^{\mu}_{\rho_I\ra\pi\pi} \frac{1}{s_{\rho}}
\Pi_{\rho\omega}
\frac{1}{s_{\omega}} \M_{\gamma\ra\omega_I},
\end{equation}
where $1/s_V$ is the vector meson propagator.

The couplings that enter this expression, through
$\M^{\mu}_{\rho_I\ra\pi\pi}$, $\M_{\gamma\ra\rho_I}$ and
$\M_{\gamma\ra\omega_I}$, involve the pure isospin
states
$\rho_I$ and $\omega_I$.  However, we can re-express it in
terms of the physical states by first diagonalising the vector meson
propagator. This leads to the result
\begin{eqnarray}
\nonumber
\M^{\mu}_{\gamma\ra\pi\pi} &=&
\M^{\mu}_{\rho\ra\pi\pi} \frac{1}{s_{\rho}} \M_{\gamma\ra\rho}
 + \M^{\mu}_{\omega\ra\pi\pi} \frac{1}{s_{\omega}}
\M_{\gamma\ra\omega} \\
&=&
\M^{\mu}_{\rho\ra\pi\pi} \frac{1}{s_{\rho}}
\M_{\gamma\ra\rho}
+ \M^{\mu}_{\rho\ra\pi\pi}
\frac{\Pi_{\rho\omega}}{s_\rho-s_\omega}
\frac{1}{s_{\omega}} \M_{\gamma\ra\omega} ,
\end{eqnarray}
which is the form usually seen in older works \cite{CSM,CB,MSC,us}.
A recent analysis \cite{Heath} of the world data gave a value for the
mixing amplitude of $\Pi^{\rho \omega} = -3800 \pm 370$ MeV$^2$.

\section{Recent Tests and Applications}\label{sec:recent}
\subsection{Testing CVC in $\tau^-$ Decay}\label{subsec:tau}
All modern $e^+e^-$ colliders with sufficient energy (including LEP)
have studied the decay
\begin{equation}
\tau^- \ra \pi^- + \pi^0 + \nu_{\tau}.
\end{equation}
At present the world average for the branching ratio into this
channel is\cite{pdg} $25.24 \pm 0.16$\%.  Only the hadronic vector
current is involved and according to CVC this will be only the I=1
piece of the vector current.  Thus, unlike $e^+ e^- \ra \pi^+
\pi^-$, there will be no $\rho - \omega$ interference.

The procedure is therefore to fit the $e^+ e^- \ra \pi^+ \pi-
$ data, including $\rho - \omega$ mixing as well as the $\rho'
(1450)$ and $\rho''(1700)$, and then use the purely I=1 piece of
the amplitude to calculate the decay rate \cite{Sobie,EI}.  The most
recent estimate of the theoretical branching ratio comes from the
analysis of Sobie\cite{Sobie}, namely $24.6 \pm 1.4$\%.  This differs
from experiment by about $2\frac{1}{2}$\% with a mainly theoretical
error of $\sim 5$\%.  Once again CVC works but the level of accuracy
is relatively modest.  In order to improve the accuracy one would
need to resolve the differences between the existing data sets for
$e^+ e^- \ra \pi^+ \pi^-$.

\subsection{An Application to CP Violation}\label{subsec:CP}
The fact that CVC is not exact, and in particular the mixing of
$\rho$ and $\omega$, can be put to use in quite a spectacular fashion
in the study of CP violation.  We shall briefly report on the recent
analysis of Enomoto and Tanabashi\cite{ET}, following a suggestion of
Lipkin\cite{Lip}.  Rather than the traditional proposals to study
$(B^0,\bar{B}^0)$ decays, these authors aim to {\it use} $\rho -
\omega$ interference to generate a large CP violation signal in
charged $B$ decays.

One can show that the CP-violating difference in the decay rates
$B^- \ra \rho^0 \rho^-$ and $B^+ \ra \rho^0 \rho^+$ is proportional
to $cos(\delta + \phi) - cos (\delta - \phi)$, where $\phi$ is the
CP violation phase and $\delta$ the known, strong phase arising from
$\rho - \omega$ mixing.  Notice that the signal vanishes if
$\delta = 0$.  Although the branching ratios for the decay modes
$B^- \ra \rho^0 h^-$ (with $h = K, K^*,\rho$, etc.) is very small
($\sim 10^{-8}$), the asymmetry can be as large as 90\% !  This is
clearly a very important suggestion to pursue further.

In concluding this brief summary we note one question which needs
urgent study.  At recent analysis of $\rho - \omega$ mixing by
Maltman {\it et al.}\cite{Malt} has suggested that, contrary to
earlier conclusions, the direct coupling of the I = 0 $\omega$ to
two pions leads to a large uncertainty in $\Pi^{\rho \omega}$ and
in the relative phase.  In view of the need to know the strong phase
$\delta$ very well, in order to extract $\phi$, this is a worrying
conclusion.  We simply note that this ambiguity vanishes identically
at $q^2 = \hat{m}^2_{\rho}$, but that the residual ambiguity needs
careful study.

\section{Beyond the Standard Model}\label{sec:sm}
The Standard Model has proven very successful in
every area of particle physics, including recent 
high-energy collider experiments.
However, it has three features which are not well understood:
the origin of mass, the three fermion generations and the phenomenon of
CP violation.
The question of mass is usually framed in terms of (fundamental) Higgs
fields \cite{Higgs} and why
the corresponding Yukawa couplings take particular values. 
Instead, one might ask
whether a formulation of the Standard Model with massless fermions 
makes sense. For example, it is well known that QED
with massless electrons
is not well defined at the quantum level \cite{Muta,8}.

In a recent paper, Bass and Thomas \cite{BT} considered 
the pure Standard Model with gauge symmetry
$SU(3) \otimes SU(2)_L \otimes U(1)$ and no additional interaction --
i.e., with no grand unification.
They examined the physical theory corresponding to the bare Standard
Model Lagrangian
with no elementary Higgs and just one generation of massless 
fermions and gauge
bosons. 
At asymptotic scales, where the U(1) coupling is significantly
greater
than
the asymptotically free SU(3) and SU(2)$_L$ couplings,
the left and right handed states of any given charged fermion couple to
the U(1) gauge boson
with different charges.
At the Landau scale of this non-asymptotically free theory, it was
suggested that there should be three separate
phase transitions -- corresponding to each of the right-right, right-left
and left-left interactions becoming supercritical. These transitions
correspond to three generations of fermions. As one passes through each
transition from a higher scale (shorter distance) the corresponding
scalar condensate ``melts'', releasing a dynamical fermion into the
Dirac phase studied in the laboratory. In this picture the three
generations emerge as quasi-particle states built on a ``fundamental
fermion'' interacting self-consistently with the condensates.

Clearly this proposal differs in a fundamental manner from the
conventional approaches to the Standard Model. While the conceptual
framework is extremely simple and 
elegant, the techniques for dealing with
non-perturbative physics at the Landau scale are not well developed.
In particular, at the present stage it has not yet been possible to present a
rigorous, quantitative derivation of all of the features of the Standard
Model. Nevertheless, we believe that the potential for understanding so
many phenomena, including mass, CP-violation and the generations, not to
mention CVC, is so
compelling that the ideas merit further study. 

\section{Conclusion}\label{sec:conc}
We have seen clearly that CVC is a very natural approximation
within the Standard Model.  Certainly the isovector, vector current
(modulo the CKM matrix) is exactly the vector current involved in
the charged current weak interactions.  The hypothesis that the
current is conserved is an approximation only because $m_u \neq m_d$
-- and, of course, because of the electromagnetic interaction.

Within particle physics the best test of CVC is pion $\beta$-decay,
with the present limit being about 3\%.  Using the decay of the
heavier (1.777 GeV) $\tau^-$ to $\pi^0 \pi^- \nu_{\tau}$ one
can set a limit that is only slightly worse, around 5\%. Tests
involving neutrino-nucleon scattering are quite imprecise, with
the fundamental Adler sum-rule (which, of course, tests current
algebra not just CVC) being in desperate need of accurate data.

We also saw that $\rho - \omega$ mixing, which is a spectacular
example of the non-conservation of the vector current, provides a
very beautiful alternative way to study CP violation in $B$-decays. 
Finally, we briefly reviewed a rather ambitious framework for
understanding the origin of many phenomenological features of the
Standard Model, including the three generations and their masses.
Such an approach may someday provide us with a real understanding
of the origins of CVC.

\section{Acknowledgements}\label{sec:Ack}
I would like to thank a number of colleagues for helpful discussions,
especially H. O'Connell and A.G. Williams concerning
$\rho - \omega$ mixing, S.V. Gardner concerning the CP violation
proposal and S.D. Bass concerning our generations proposal.  It is
a pleasure to thank Professor Minamisono and his colleagues for the
opportunity to participate in a most stimulating symposium.  This work
was supported by the Australian Research Council.


%
\end{document}